\documentclass[prl, aps, reprint, superscriptaddress,showpacs,amsmath,amssymb, twocolumn, floatfix]{revtex4-1}

\usepackage{graphicx}
\usepackage{dcolumn}
\usepackage{bm}
\usepackage{color}
\usepackage{ulem}
\usepackage[utf8]{inputenc}
\usepackage{siunitx}
\usepackage{hyperref}

\newcommand{\Omegam}{\Omega_{\mathrm{m}}}
\newcommand{\omegac}{\omega_{\mathrm{c}}}
\newcommand{\Phib}{\Phi_{\mathrm{b}}}
\newcommand{\phib}{\phi_{\mathrm{b}}}

\begin{document}


\title{Mechanical frequency control in inductively\\coupled electromechanical systems}

\author{Thomas Luschmann}
\email[]{thomas.luschmann@wmi.badw.de}
\affiliation{Walther-Meißner-Institut, Bayerische Akademie der Wissenschaften, Walther-Meißner-Str.8, 85748 Garching, Germany}
\affiliation{Physik-Department, Technische Universität München, James-Franck-Str.1, 85748 Garching, Germany}
\affiliation{Munich Center for Quantum Science and Technology, Schellingstr.4, 80799, Munich, Germany}

\author{Philip Schmidt}
\altaffiliation[Present address: ]{Institute for Quantum Optics and Quantum
Information, Austrian Academy of Sciences, 1090 Vienna, Austria}
\affiliation{Walther-Meißner-Institut, Bayerische Akademie der Wissenschaften, Walther-Meißner-Str.8, 85748 Garching, Germany}
\affiliation{Physik-Department, Technische Universität München, James-Franck-Str.1, 85748 Garching, Germany}

\author{Frank Deppe}
\affiliation{Walther-Meißner-Institut, Bayerische Akademie der Wissenschaften, Walther-Meißner-Str.8, 85748 Garching, Germany}
\affiliation{Physik-Department, Technische Universität München, James-Franck-Str.1, 85748 Garching, Germany}
\affiliation{Munich Center for Quantum Science and Technology, Schellingstr.4, 80799, Munich, Germany}

\author{Achim Marx}
\affiliation{Walther-Meißner-Institut, Bayerische Akademie der Wissenschaften, Walther-Meißner-Str.8, 85748 Garching, Germany}

\author{Alvaro Sanchez}
\affiliation{Department of Physics, Universitat Autonoma de Barcelona, 08193 Bellaterra, Catalonia, Spain}

\author{Rudolf Gross}
\affiliation{Walther-Meißner-Institut, Bayerische Akademie der Wissenschaften, Walther-Meißner-Str.8, 85748 Garching, Germany}
\affiliation{Physik-Department, Technische Universität München, James-Franck-Str.1, 85748 Garching, Germany}
\affiliation{Munich Center for Quantum Science and Technology, Schellingstr.4, 80799, Munich, Germany}

\author{Hans Huebl}
\email[]{hans.huebl@wmi.badw.de}
\affiliation{Walther-Meißner-Institut, Bayerische Akademie der Wissenschaften, Walther-Meißner-Str.8, 85748 Garching, Germany}
\affiliation{Physik-Department, Technische Universität München, James-Franck-Str.1, 85748 Garching, Germany}
\affiliation{Munich Center for Quantum Science and Technology, Schellingstr.4, 80799, Munich, Germany}

\date{\today}

\begin{abstract}
Nano-electromechanical systems implement the opto-mechanical interaction combining electromagnetic circuits and mechanical elements. We investigate an inductively coupled nano-electromechanical system, where a superconducting quantum interference device (SQUID) realizes the coupling. We show that the resonance frequency of the mechanically compliant string embedded into the SQUID loop can be controlled in two different ways: (i) the bias magnetic flux applied perpendicular to the SQUID loop, (ii) the magnitude of the in-plane bias magnetic field contributing to the nano-electromechanical coupling. These findings are quantitatively explained by the inductive interaction contributing to the effective spring constant of the mechanical resonator. In addition, we observe a residual field dependent shift of the mechanical resonance frequency, which we attribute to the finite flux pinning of vortices trapped in the magnetic field biased nanostring. 
\end{abstract}
\maketitle
The opto-mechanical interaction couples the displacement of a mechanical mode to the resonance frequency of an optical resonator. This setting is key for the realization of ultra-sensitive force detectors and the investigation of quantum mechanics in the literal sense \cite{abbott2016,AspelmeyerRMP,chan2011,delic2020}. One sub-field of common opto-mechanical systems is the field of circuit- or nano- electromechanical systems, where the role of the optical resonator, operating in the terahertz range, is taken over by a gigahertz microwave resonator. For the majority of devices, the opto-mechanical interaction is realized by capacitive coupling, where the displacement of the mechanical element is transduced into a variation of the microwave resonator capacitance, thereby shifting its resonance frequency \cite{regal2008,teufel2011,zhou2013}. Here, vacuum coupling rates have reached the widely reported limit of $\SI{300}{Hz}$ \cite{blencowe2007,buks2007,nation2016,Reed2017} and effects such as ground state cooling \cite{teufel2011a}, strong coupling \cite{teufel2011,peterson2019}, state transfer \cite{palomaki2013}, mechanical squeezing \cite{wollman2015}, as well as electromechanically induced transparency effects \cite{zhou2013,singh2014,hocke2012,weber2016} have been reported. 
Only recently, the concept of inductive coupling, where the mechanical displacement is transduced into a change of the resonator inductance and hence resonance frequency, has been demonstrated \cite{rodrigues2019,schmidt2020,zoepfl2020,bera2021}. This concept offers the potential to significantly increase the single-photon coupling rate $g_0$ beyond the above-mentioned limit \cite{blencowe2007,buks2007,nation2016,shevchuk2017,Reed2017}. Devices realizing this concept are based on a direct-current superconducting quantum interference device (dc-SQUID), incorporated into a coplanar waveguide (CPW) microwave resonator. Here, the dc-SQUID can be viewed as a nonlinear, flux-controllable inductor. The integration of a mechanically compliant system into one arm of the dc-SQUID enables the opto-mechanical interaction. Notably, the mechanical element  experiences a backaction fundamentally based on the Lorentz force \cite{poot2010,shevchuk2017,etaki2008}, which has so far only been experimentally explored in isolated dc-SQUID based electromechanical devices operated in the voltage mode. This raises several interesting questions: (i) Is this effect present in circuit-integrated nano-electromechanical devices, (ii) does it represent an efficient way to tune the mechanical frequency, and (iii) do the magnetic fields controlling the optomechanical interaction constant $g_0$ affect the mechanical properties of the device. While indications of the frequency tuning effect have been briefly mentioned in a recent publication \cite{rodrigues2019}, a detailed study of the effect regarding the above questions is still missing.

With this motivation, we present experimental data of a nano-electromechanical system based on an inductive coupling scheme, where we explore the tuning of the mechanical resonance frequency as a function of multiple control parameters. In the experiment, we observe a flux-dependent shift of the mechanical resonance frequency, which we can quantitatively describe by theoretical predictions. Interestingly, we observe an additional, flux-independent shift of the mechanical frequency which we attribute to the influence of magnetic flux lines trapped in the superconducting nanostring.

\begin{figure}
	\includegraphics[scale=1]{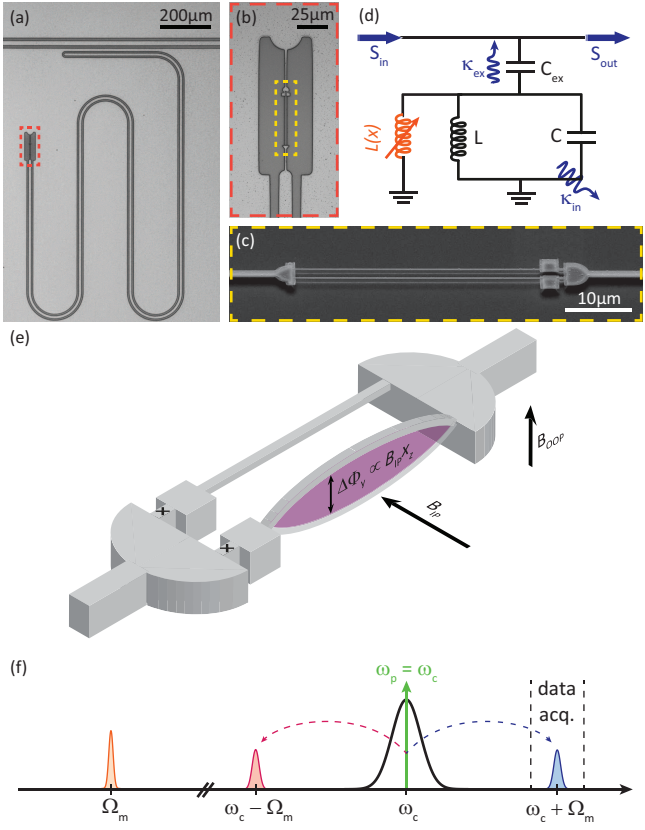}
	\caption{(a) optical micrograph image of the $\lambda/4$ coplanar-waveguide resonator coupled to a feedline (top) and short-circuited to ground via a flux-dependent inductance formed by a dc-SQUID (orange box). (b) Magnified view of the dc-SQUID with freely suspended strings. (c) Tilted scanning electron micrograph image of a suspended SQUID structure similar to the one used in this work. Note that the actual device features nanostrings of $\SI{20}{\mu m}$ length. (d) Equivalent circuit representation of the device. The CPW resonator is described by means of an effective capacitance $C$ and inductance $L$, forming an LC-oscillator. Additionally, the circuit contains a dynamic inductance (orange), whose magnitude depends on the time-varying displacement of the mechanical element. (e) Illustration of the SQUID with incorporated nanostrings and relevant magnetic field directions. The shading illustrates how the motion of the nanostring modulates the flux-threaded area of the SQUID-loop due to an in-plane magnetic field $B_\mathrm{IP}$. (f) Overview of the relevant frequencies and microwave tones. For the determination of the mechanical resonance frequency $\Omegam$ we perform spectral analysis of the anti-Stokes field.}
	\label{fig:one}
\end{figure}
The device investigated in this letter is shown in Fig.~\ref{fig:one} (a)-(c). A $\lambda/4$ superconducting CPW resonator is short-circuited to ground via a dc-SQUID at one end (cf. Fig. \ref{fig:one}b). The inductance of the SQUID is flux-dependent, allowing for the control of the resonance frequency via an out-of-plane oriented applied magnetic field $B_\mathrm{OOP}$. To enable the electromechanical interaction, parts of the SQUID loop are suspended, forming two nanomechanical string oscillators (cf. Fig. \ref{fig:one}c). The displacement of the strings modulates the effective area of the SQUID and therefore alters the magnetic flux threading the loop (cf. Fig. \ref{fig:one}e). Since the inductance of the dc-SQUID depends on the applied flux, this in turn results in a modulation of the resonance frequency of the microwave resonator. We note that the nanostring oscillators support both out-of-plane (OOP) and in-plane (IP) flexural modes. By applying a magnetic field parallel (perpendicular) to the chip plane, the circuit becomes sensitive to the OOP (IP) displacement of the nanostrings. In this experiment, we use a strong in-plane magnetic field $B_\mathrm{IP}$ to realize an enhanced electromechanical interaction, since the in-plane field orientation supports a much higher critical magnetic field compared to the OOP direction \cite{meservey1971}. Additionally, we employ a weak magnetic field $B_\mathrm{OOP}$ to control the resonance frequency of the microwave resonator.

According to Ref.~\onlinecite{shevchuk2017}, the system can be described in terms of the mechanical displacement $X$ and the center-of-mass coordinate $\varphi_+ = (\phi_1+\phi_2)/2$ of the SQUID, where $\phi_1$ and $\phi_2$ are the phase differences across the Josephson junctions. The resulting Hamiltonian can be written as 
\begin{equation}
   H = \frac{m_\mathrm{r} \dot{X}^2}{2}+\frac{m_r \Omegam^2 X^2}{2}+ \frac{C \Phi_0^2}{2(2\pi)^2}\dot{\varphi_+}^2+E(\varphi_+,X).
    \label{eq:shevchuk_hamiltonian}
\end{equation}
Here, $m_\mathrm{r}$ and $\Omegam$ are the mass and the resonance frequency of the mechanical string, $\Phi_0$ the flux quantum and $E(\varphi_+,X)$ represents the potential energy of the dc-SQUID. As the resonance frequency of the microwave cavity, $\omega_c$, is much larger than the mechanical resonance frequency ($\omegac \gg \Omegam$), the system operates in the dispersive limit and the mechanical displacement along with the corresponding flux change can be considered as static on the timescales relevant for the SQUID dynamics. In this limit, the SQUID can be approximated as a harmonic oscillator and an expansion of its potential energy in terms of the phase up to second order can be performed. The resulting phase-dependent terms describe the dynamic interaction of the electromechanical system, which eventually lead to a radiation pressure interaction term of the form $H_\mathrm{int} = \hbar g_0 \hat{a}^\dagger \hat{a}(\hat{b}^\dagger + \hat{b})$, where $\hat{a}$ and $\hat{b}$ denote the ladder operators of the microwave resonator and mechanical oscillator, respectively \cite{schmidt2020,rodrigues2019,blencowe2007,buks2007,nation2016}. To investigate the mechanical resonance frequency, we instead focus on the phase-independent term in the expansion, which induces a static, but flux-dependent shift of the uncoupled mechanical resonance frequency $\Omega_0$ towards an effective frequency $\Omegam$. According to \cite{shevchuk2017}, one obtains
\begin{equation}
   \Omegam =\sqrt{\Omega_0^2 +\frac{4E_\mathrm{J}\pi^2B_\mathrm{IP}^2l^2\lambda^2(1-\alpha^2)[\cos^4(\phib)-\alpha^2\sin^4(\phib)]}{m_\mathrm{r} S_0^3}}.
    \label{eq:shevchuk_freq}
\end{equation}
Here, $E_\mathrm{J} = \hbar(I_1+I_2)/4e$ is the SQUID average Josephson energy with $I_{1,2}$ representing the critical currents of the individual Josephson junctions, $l$ is the length of the string and $\lambda$ its shape factor \cite{poot2010}. Furthermore, we have introduced $S_0=\sqrt{\cos^2(\phib)+\alpha^2\sin^2(\phib)}$ with the normalized bias flux $\phi_\mathrm{b} = \pi\Phib/\Phi_0$, which is generated by the control field $B_\mathrm{OOP}$. The asymmetry parameter $0 < \alpha < 1$ accounts for non-identical Josephson junctions in the SQUID and is defined by $I_1 = I_0(1-\alpha)$ and $I_2 = I_0(1+\alpha)$ with the average critical current $I_0 = (I_1 + I_2)/2$.

The modification of the mechanical resonance frequency reflected in Eq.\,(\ref{eq:shevchuk_freq}) can be understood as a consequence of a total Lorentz force $F_\mathrm{L} = B_\mathrm{IP}lI(\Phi)$ originating from the flux-dependent circulating current $I(\Phi)$ of the dc-SQUID. However, as the total flux $\Phi$ includes $\phib$, which is modulated by the mechanical motion of the nanostring, a displacement-dependent restoring force is obtained. This corresponds to an effective change in the strings stiffness and hence a modification of the mechanical resonance frequency. Note that this type of backaction is different from the shift in $\Omegam$ caused by the opto-mechanical interaction \cite{AspelmeyerRMP,teufel2008}.

The device depicted in Fig.\,\ref{fig:one} (a)-(c) is fabricated using standard nanofabrication techniques, double-layer shadow evaporation of aluminium and reactive ion etching on high resistivity silicon (see Ref.~\onlinecite{schmidt2020} for the fabrication details of this particular device). The suspended nanostrings have dimensions of $(l,w,t) = (20,0.2,0.11)\si{\mu m}$, resulting in an effective mass of $m_\mathrm{r} =\SI{0.6}{pg}$ and mechanical out-of-plane frequencies of $\Omegam/2\pi \approx \SI{5.8}{MHz}$ at millikelvin temperatures.

\begin{figure}
	\includegraphics[scale=1]{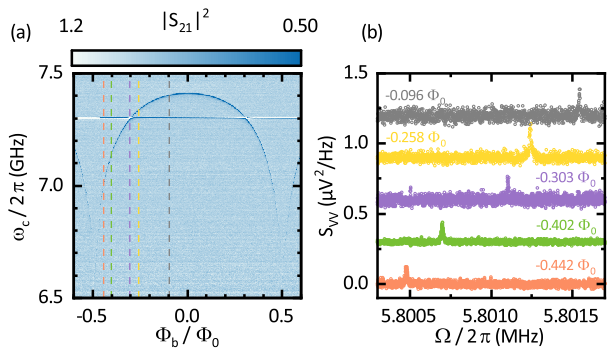}
	\caption{(a) Calibrated microwave transmission magnitude as a function of the normalized applied magnetic flux. The flux-dependent inductance of the circuit allows for the adjustment of the resonance frequency from \SIrange{7.45}{6.6}{GHz}. A parasitic resonance is visible around \SI{7.3}{GHz}. Colored dashed reference lines correspond to the flux bias points at which the mechanical resonance frequency is analyzed in detail.  (b) Voltage power spectral density of the demodulated probe tone, analyzed around its anti-Stokes peak at various flux bias points (see labels next to the data trace) for a fixed in-plane field of $B_\mathrm{IP} = \SI{35}{mT}$. The spectra are offset by $\SI{0.3}{\mu V^2/Hz}$ each for clarity. The Lorentz-shaped mechanical resonance features shift to higher frequencies by roughly \SI{1}{kHz} as the normalized flux bias is decreased. }
	\label{fig:two}
\end{figure}
All experiments are performed in a dilution refrigerator at a temperature of approximately $\SI{85}{mK}$ using spectroscopy schemes such as microwave transmission experiments and thermal sideband noise spectroscopy. A detailed description of the microwave detection setup can be found in Ref.~\onlinecite{schmidt2020}. To apply the strong in-plane and the weak out-of-plane magnetic field to the nano-electromechanical circuit, we position the chip in a superconducting solenoid magnet and mount a small superconducting coil on the sample enclosure. While the IP superconducting solenoid is used to apply fields of up to \SI{35}{mT}, the field provided by the OOP coil is limited to \SI{1}{mT}. In order to achieve undisturbed exposure of the device to $B_\mathrm{IP}$, we operate without magnetic shielding. As a consequence, the device is subject to fluctuations in the static magnetic field which are actively compensated by a feedback loop using the OOP coil as control entity. 

We start with the characterization of the flux-tunable CPW resonator using microwave transmission measurements. We record the frequency dependent complex transmission using a vector network analyzer for various OOP flux bias values $\Phib$. The results are shown in Fig.\,\ref{fig:two}(a) for $B_\mathrm{IP} = 0$. The resonance frequency of the microwave resonator, $\omegac$, is visible as a dark blue feature in the color-coded scattering parameter $|S_{21}|^2$. We observe a maximum frequency of $\omegac/2\pi\approx \SI{7.45}{GHz}$, which decreases as $\Phib$ is increased due to the increasing Josephson inductance of the SQUID. This effect is periodic in $\Phib$ as expected for such resonators \cite{sandberg2008,schmidt2020, rodrigues2019,pogorzalek2017} and allows for the experimental control of the resonance frequency over a range of roughly $\SI{750}{MHz}$. In zero-(OOP)-field we find a total linewidth of $\kappa/2\pi \approx \SI{2.5}{MHz}$ and a minimum Josephson inductance of the SQUID $L_\mathrm{J} = \SI{0.36}{nH}$. A more detailed analysis of this device is presented in Ref.~\onlinecite{schmidt2020}.

In order to investigate the impact of the electromechanical system on the mechanical subsystem's frequency, we analyze the properties of the mechanical resonator as function of $\Phib$ and the in-plane bias field $B_\mathrm{IP}$. To this end, we record a thermal displacement spectrum of the anti-Stokes field as illustrated in Fig.\,\ref{fig:one}(f). In detail, we inject a weak probe tone, which is resonant with the microwave resonator [$\omega_\mathrm{p} = \omegac(\Phib)$]. In addition, a second, weaker stabilizer tone is applied at $\omega_\mathrm{stab} = \omegac(\Phib)+\SI{500}{kHz}$ (not shown). The transmission of the stabilizer tone constitutes the error signal input used for the active feedback provided via $B_\mathrm{OOP}$ to counteract any magnetic field fluctuations. We use tones with small detuning and ultra-low powers ($P_\mathrm{stab} < P_\mathrm{probe} < \SI{2}{fW}$) to avoid opto-mechanical heating and cooling effects as well as the resulting frequency shift (opto-mechanical spring effect) \cite{schmidt2020,teufel2008,AspelmeyerRMP}.

The thermal displacement noise of the nanostring modulates the inductance of the SQUID and thus $\omegac$. The resulting sidebands, which correspond to the Stokes and anti-Stokes field, appear at $\omega_\mathrm{p} \pm \Omegam(\Phi_\mathrm{b})$. We record the spectral density of the anti-Stokes field \footnote{We use $\omega_c(\Phi_\mathrm{b}) + \SI{3.5}{MHz}$ as downconversion frequency and selectively analyze the anti-Stokes field} and fit a Lorentzian lineshape to the data to characterize the mechanical oscillator. We extract the mechanical resonance frequency $\Omegam/2\pi \approx \SI{5.8}{MHz}$ and the linewidth $\Gamma_\mathrm{m}/2\pi \approx \SI{20}{Hz}$, corresponding to $Q \approx \num{290000}$. Based on Eq.\,(\ref{eq:shevchuk_freq}) we expect an evolution of $\Omegam$ with $B_\mathrm{OOP}$ or flux bias $\Phib$. Fig.\,\ref{fig:two}(b) shows the recorded voltage power spectral density for various flux bias points and a fixed $B_\mathrm{IP}$ of $\SI{35}{mT}$. Similar to the behavior of the microwave resonator, the mechanical frequency appears to decrease with an increasing flux bias applied to the SQUID.

\begin{figure}
	\includegraphics[scale=1]{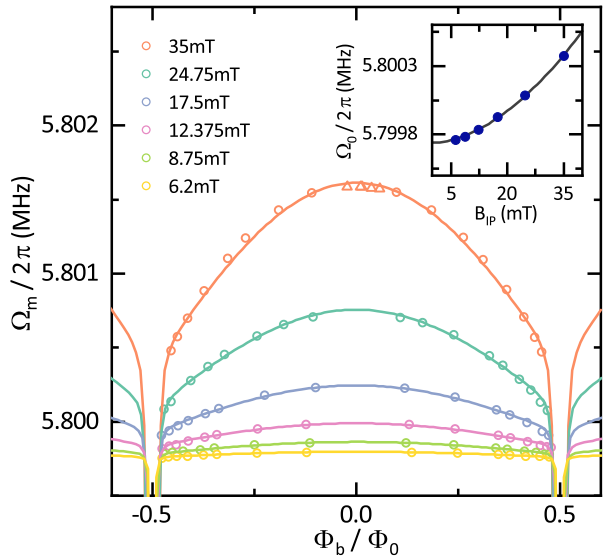}
	\caption{Extracted mechanical resonance frequency as a function of the applied bias flux through the SQUID loop, measured at different in-plane fields $B_{\mathrm{IP}}$ ranging from \SIrange{6.2}{35}{mT}. Open circles correspond to data gathered by measurement of the thermal motion, while triangular data points are acquired with a piezoelectric actuator resonantly driving the mechanical motion. Lines are fits to the data according to Eq.\,(\ref{eq:shevchuk_freq}) with $E_\mathrm{J}$ and $\Omega_\mathrm{0}$ as the only free fit parameters. The inset shows $\Omega_\mathrm{0}$ as a function of the applied in-plane field as extracted from the fits as well as a power-law fit (black line) revealing $\Omega_\mathrm{0} \propto B_\mathrm{IP}^{1.81}$. The increase suggests an additional contribution not included in the presented model and is discussed in the main text. Statistical error bars are smaller than the symbol size.}
	\label{fig:three}
\end{figure}

To gain a deeper understanding of the effects at play, we perform the previously described experiment at various flux bias points $\Phib$ covering the full periodicity of the microwave resonator frequency. In addition, we repeat these flux sweeps for various in-plane magnetic fields ranging from \SIrange{6.2}{35}{mT}. The extracted mechanical resonance frequencies are plotted in Fig.\,\ref{fig:three}. We note that the electromechanical coupling is strongly suppressed at small flux bias ($|\Phi_\mathrm{b}/\Phi_0| < 0.1$). Therefore, to verify the data in this regime, a piezoelectric actuator attached to the sample was used to resonantly drive the mechanical motion and increase the signal strength. 

As shown in Fig~\ref{fig:three}, the mechanical resonance frequency $\Omegam$ approximately shows a parabolic tuning behavior with respect to the applied flux bias. We observe a maximum tuning of roughly \SI{1}{kHz} at the maximum applied in-plane field, $B_\mathrm{IP} = \SI{35}{mT}$. We compare these results with the theoretical prediction by fitting the data to Eq.\,(\ref{eq:shevchuk_freq}), choosing $E_\mathrm{J}$ and $\Omega_0$ as the only free fit parameters. The SQUID asymmetry is fixed to a small value of $\alpha = 1\%$ to account for minor fabrication deviations of the Josephson junctions, while the remainder of the device parameters has been determined in previous experiments \cite{schmidt2020} and are summarized in the supplement \cite{SupplementaryInformation}.
The resulting model precisely describes our experimental findings, confirming the hypothesis that the shift of the mechanical resonance frequency is indeed caused by the Lorentz force acting on the nanostring within the SQUID.
However, an analysis of the determined fit parameters (see inset) reveals that the uncoupled resonance frequency $\Omega_\mathrm{0}$ increases by several hundred \si{Hz} as $B_\mathrm{IP}$ is increased, an effect that is not accounted for by our theoretical description of the electromechanical system [cf. Eq.\,(\ref{eq:shevchuk_freq})]. Since the shift in $\Omega_0$ is independent of $\Phib$, we cannot attribute it to the common first-order opto-mechnical interaction.

Next, we discuss various scenarios which could explain the $B_\mathrm{IP}$-dependence of $\Omega_0$. One possible mechanism could be the volume change of the superconducting aluminium as the in-plane field approaches its critical field \cite{ott1972}. However, the commonly reported length changes on the order of $\Delta V/V \approx 10^{-8}$ are too small to explain the observed frequency shifts $\Delta\Omegam/\Omegam \approx 10^{-4}$. Higher order contributions to the opto-mechanical interaction, specifically the quadratic opto-mechanical coupling, would result in a $B_\mathrm{IP}^2$-dependence of $\Omegam$ \cite{liao2015}. However, for our device parameters, we expect a frequency shift on the order of a few \si{Hz} and hence refrain from this conjecture. We also consider that the flux captured by the SQUID loop and the corresponding magnetic moment could give rise to a modification of the mechanical frequency if the device acts as a torque magnetometer \cite{kamra2015,kamra2014,petkovic2020,misakian2000}. However, controlling the number of flux quanta in the SQUID loop via $B_\mathrm{OOP}$ allows us to rule out this conjecture (for details see supplement \cite{SupplementaryInformation}).

Finally, we draw parallels to vibrating reed experiments, which were able to measure the stiffness of the flux line lattice (FLL) in type-II superconductors and its influence on mechanical properties \cite{brandt1986,esquinazi1991}. In particular, the FLL can exhibit quasi-elastic properties, couple to the motion of the atomic lattice and hence influence the mechanical resonance frequency. The strength of this coupling is quantified by the Labusch parameter $\alpha_\mathrm{L}(B,T)$. Notably, in the case of very thin superconductors, as used for our experiment, the expected frequency change is $\Omega^2 = \Omega_0^2 + \alpha_L(B)/\rho$ \cite{esquinazi1991}.
The field-dependence of $\alpha_\mathrm{L}(B)$ exhibits a power-law behavior $\alpha_\mathrm{L}(B) \propto B^k$ where $k$ can vary significantly across materials. While we are not aware that this effect has been reported for aluminium, sufficiently thin aluminium films can behave in ways characteristic of type-II superconductors \cite{brandt1971,khukhareva1963}. Studies of other type-II superconductors find a range of $k \approx 2 \pm 0.5$ and $\alpha_\mathrm{L} \approx \num{d12}$ to $\SI{d15}{N/m^4}$ \cite{esquinazi1991,gupta1991,kober1991}. Fitting our experimental data with a power-law (Fig. \ref{fig:three} inset) we find $\alpha_\mathrm{L} \propto B_\mathrm{IP}^{1.81}$ and $\alpha_\mathrm{L} (\SI{35}{mT})= \SI{7.88d14}{N/m^4}$, showing agreement with the reported values. While this suggests that flux line pinning is at the origin of the mechanical frequency shift, further investigations are warranted to fully confirm this hypothesis.

In summary, we present a detailed study of frequency tuning effects on a nanostring in an electromechanical system. We find a pronounced shift ($> 50\, \Gamma_\mathrm{m}$) in the mechanical resonance frequency as function of the flux bias condition of the SQUID. This $\Phib$-dependent frequency shift is distinct from frequency shifts based on the opto-mechanical interaction and is quantitatively explained by an interaction induced by the SQUID via the Lorentz force. Furthermore, our detailed modelling reveals a previously unobserved field-dependent frequency shift that we attribute to the additional mechanical stiffness induced by the flux line lattice in the aluminium nanostring. This underlines the mechanical sensing capabilities of nanostrings and their potential applications in material science. The reduced size of this particular system promises a novel approach to the investigation of the mechanical properties of few or individual flux lines. The detailed understanding of the frequency shifts, as it is presented in this work, is of utmost importance for applications relying on the precise in-situ control of the mechanical resonator frequency.
\begin{acknowledgments}
This project has received funding from the European Union's Horizon 2020 research and innovation program under grant agreement No 736943 and from the Deutsche Forschungsgemeinschaft (DFG, German Research Foundation) under Germany’s Excellence Strategy—EXC-2111-390814868. A.S. acknowledges funding from ICREA Academia, Generalitat de Catalunya. We gratefully acknowledge valuable scientific discussions with D. Koelle.
\end{acknowledgments}
%

\end{document}